\documentclass[pra,twocolumn,aps,letterpaper, preprintnumbers,superscriptaddress]{revtex4-2}

\usepackage{amsthm,amsmath,amssymb}

\usepackage{mathrsfs}
\usepackage{amsthm}
\usepackage{amsmath}
\usepackage{amssymb}
\usepackage{graphicx}
\usepackage{epstopdf}
\usepackage{url}
\usepackage{subfigure}
\usepackage{float}
\usepackage{bm}
\usepackage{ifthen}
\usepackage[usenames,dvipsnames]{color}
\usepackage{mathrsfs}
\usepackage[bookmarks=true,colorlinks, citecolor=blue, linkcolor=blue, urlcolor=blue]{hyperref}
\usepackage[dvipsnames, svgnames, x11names]{xcolor} 
\usepackage{braket}
\usepackage{color}
\usepackage{indentfirst}

\theoremstyle{plain}
\newtheorem{theorem}{Theorem}[section]

\theoremstyle{definition}
\newtheorem{definition}{Definition}[section]

\theoremstyle{remark}

\begin{document}

\date{\today}

\title{Asynchronous measurement-device-independent quantum digital signatures}

\author{Jing-Wei Bian}
\author{Bing-Hong Li}	
\author{Yuan-Mei Xie}
\affiliation{National Laboratory of Solid State Microstructures and School of Physics, Collaborative Innovation Center of Advanced Microstructures, Nanjing University, Nanjing 210093, China}
\affiliation{Department of Physics and Beijing Key Laboratory of Opto-electronic Functional Materials
and Micro-nano Devices, Key Laboratory of Quantum State Construction and Manipulation (Ministry of Education),
Renmin University of China, Beijing 100872, China}
\author{Hua-Lei Yin}\email{hlyin@ruc.edu.cn}
\affiliation{Department of Physics and Beijing Key Laboratory of Opto-electronic Functional Materials
and Micro-nano Devices, Key Laboratory of Quantum State Construction and Manipulation (Ministry of Education),
Renmin University of China, Beijing 100872, China}
\affiliation{National Laboratory of Solid State Microstructures and School of Physics, Collaborative Innovation Center of Advanced Microstructures, Nanjing University, Nanjing 210093, China}
\affiliation{Beijing Academy of Quantum Information Sciences, Beijing 100193, China} 
\author{Zeng-Bing Chen}\email{zbchen@nju.edu.cn}
\affiliation{National Laboratory of Solid State Microstructures and School of Physics, Collaborative Innovation Center of Advanced Microstructures, Nanjing University, Nanjing 210093, China}	 
\date{\today}

\begin{abstract}
Quantum digital signatures (QDSs), which distribute and measure quantum states by key generation protocols and then sign messages via classical data processing, are a key area of interest in quantum cryptography. However, the practical implementation of a QDS network has many challenges, including complex interference technical requirements, linear channel loss of quantum state transmission, and potential side-channel attacks on detectors. 
Here, we propose an asynchronous measurement-device-independent (MDI) QDS protocol with asynchronous two-photon interference strategy and one-time universal hashing method. The two-photon interference approach protects our protocol against all detector side-channel attacks and relaxes the difficulty of experiment implementation, while the asynchronous strategy effectively reduces the equivalent channel loss to its square root. Compared to previous MDI-QDS schemes, our protocol shows several orders of magnitude performance improvements and doubling of transmission distance when processing multi-bit messages. Our findings present an efficient and practical MDI-QDS scheme, paving the way for large-scale data processing with non-repudiation in quantum networks.

\end{abstract}

\maketitle

\section{INTRODUCTION}
Threatened by quantum attacks and the continually emerging algorithms, the security of current classical cryptographic schemes is facing challenges. This is especially true in our contemporary society where the rapid development of internet and communication technologies results in an increasing amount of data and information that needs to be collected, stored, processed, and transmitted. 
Therefore, it is necessary to develop modern cryptography to ensure the corresponding basic elements of information security: confidentiality, integrity, authenticity, and nonrepudiation~\cite{menezes2018handbook,10:Yin2022Experimental}.

Quantum technology, which is based on quantum mechanic laws, is regarded as a profoundly promising frontier in the realm of cryptography and offers a significant approach to ensuring information security~\cite{13:Xu2020Secure,83:Weng2023Beating}. 
As the most mature technology in the realm of quantum technology, quantum key distribution~\cite{11:Charles2014BB14} has undergone rapid development~\cite{20:Pirandola2020Advances,28:chen2021integrated}. 
However, it has had various security loopholes in detection~\cite{29:zhao2008quantum,30:lydersen2010hacking} until the measurement-device-independent (MDI) quantum key distribution was proposed~\cite{32:Lo2012Measurement}, which addressed all security concerns on the detection end~\cite{33:Braunstein2012Side}. Despite significant development~\cite{37:Yin2016Measurement,13:Xu2020Secure}, the key rates of most forms of MDI protocols were still constrained by the absolute repeaterless secret-key capacity~\cite{99:pirandola2017fundamental,101:Das2021Universal,102:takeoka2014fundamental}. 
Efforts have been made to break this bound~\cite{53:azuma2015all,54:lucamarini2018overcoming,55:Xie2022Breaking,56:zeng2022mode}, one of which includes an alternative variant of MDI quantum key distribution~\cite{55:Xie2022Breaking,56:zeng2022mode} called asynchronous MDI quantum key distribution. 
This variant has the ability to asynchronously pair two successful clicks over an extended pairing time, thereby establishing a two-photon Bell state. 
As a result, the secret-key capacity is broken, leading to a higher key rate and an increased distance. In addition, the asynchronous MDI scheme offers the advantage of removing the necessity for global phase tracking and phase locking. This has been confirmed through experiments that also demonstrated its superior rate and extended range~\cite{57:Zhu2023Experimental,58:Zhou2023Experimental,zhu2024field}. 

Despite the fact that combined quantum key distribution with one-time pad can ensure confidentiality against eavesdropping, technologies safeguarding the remaining three elements are more prevalent in today’s society~\cite{menezes2018handbook}.
Digital signatures, which provide the integrity, authenticity, and non-repudiation of data processing, are a suitable technique that holds broad and promising application prospects in contemporary society~\cite{59:Cao2024Experimental,schiansky2023demonstration,jing2024experimental}. 
However, widely used classical digital signature schemes provide only computational security, so unconditionally secure classical protocols have  been proposed, trying to solve the problem~\cite{9:Amiri2015Unconditionally,86:Chaum1991Unconditionally,87:Hanaoka2000UnconditionallySD}. However, they can provide information-theoretic security under only the following two circumstances. One is the existence of an authenticated broadcast channel and secure classical channels which means that more than two out of three participants are honest~\cite{98:Lamport1982Byzantine}. The other requires a trusted authority who creates and distributes keys to each participant, and this makes the protocol vulnerable to targeted attacks against the trusted authority or even to dishonesty or incompetence on the part of the trusted authority~\cite{86:Chaum1991Unconditionally,87:Hanaoka2000UnconditionallySD}. Both of these two circumstances are infeasible in the practical world. 

Unlike classical protocols, quantum digital signatures (QDSs)~\cite{15:gottesman2001quantum,16:clarke2012experimental,17:Dunjko2014Quantum,18:Collins2014Realization} are a kind of digital signature whose security relies on the secrecy and asymmetry of shared keys generated through quantum key generation protocols (KGPs)~\cite{97:Wallden2015Quantum}, without further assumptions like an authenticated broadcast channel or a trusted authority~\cite{15:gottesman2001quantum,9:Amiri2015Unconditionally,86:Chaum1991Unconditionally,87:Hanaoka2000UnconditionallySD}. As a result, they only require authenticated classical channels and insecure quantum channels to provide information-theoretic security.
First proposed in 2001~\cite{15:gottesman2001quantum}, QDS faced some impractical experimental requirements that hindered its implementation. 
However, after approximately a decade of development, these obstacles were successfully eliminated~\cite{16:clarke2012experimental,17:Dunjko2014Quantum,18:Collins2014Realization}. 
Efforts have been undertaken to eliminate the reliance on secure quantum channels~\cite{60:Yin2016Practical,61:Amiri2016Secure}, thereby triggering many achievements both theoretically~\cite{62:Puthoor2016Measurement,63:Shang2016Quantum,64:yang2017theoretically,65:Thornton2019Continuous,66:Qu2019Multi-party,67:zhang2020practical,68:Lu2021Efficient,69:Zhang2021Twin-field,70:Zhao2021Multibit,71:Weng2021Secure,72:Qin2022Quantum,73:zhang2022practical} 
and experimentally~\cite{39:Yin2017Experimental,74:Collins2016Experimental,75:Yin2017Experimental,76:roberts2017experimental,77:zhang2018proof,78:An2019Practical,79:Ding2020280-km,80:Richter2021Agile,81:Pelet2022Unconditionally}. 
However, several limitations persist across all these schemes. Protocols that employ orthogonal encoding necessitate extra symmetrization steps, leading to the need for more secure channels~\cite{61:Amiri2016Secure}. On the other hand, schemes that use non-orthogonal encoding do not depend on additional KGP channels. However, their signature rate is susceptible to the misalignment error of the quantum channel~\cite{60:Yin2016Practical,68:Lu2021Efficient,71:Weng2021Secure}. 
More importantly, these protocols can only sign one bit at a time, which results in a low signature rate when signing multi-bit documents. 
One-time universal hashing (OTUH) QDS represented an efficient change~\cite{10:Yin2022Experimental,82:Li2023One-time}, which has made significant advancement in multi-bit signatures from single-bit signatures. 
Due to the application of universal hash functions, the signature length becomes insensitive to the document volume, thus enhancing the signature rate significantly. 
This original version is efficient, but it requires perfect keys with complete secrecy. A recently proposed variant successfully resolved this problem, which reduced the requirements on perfect keys by encrypting the generator key of the hashing function~\cite{82:Li2023One-time}.

In this work, we propose a protocol named asynchronous MDI-QDS, which delves deeply into the potential of the OTUH method. 
\textcolor{black}{Our protocol is carried out with the use of the asynchronous MDI method and the OTUH method. 
In the asynchronous MDI method, two participants send pulses to a measurement node to perform single-photon interference (SPI). Then, utilizing time multiplexing, the asynchronous two-photon interference strategy matches two successful SPI events in different time bins that are phase-correlated to obtain an asynchronous two-photon Bell state, and then, the key rate is enhanced to $O(\sqrt{\eta})$ scaling, where $\eta$ is the total channel transmittance between the two participants. This leads to a significant enhancement in the signature rates and an extension of the signature distance. 
In the OTUH method, the signature is generated by the hash function described in Appendix \ref{Appendix1} operating on the multibit documents. Compared to single-bit QDS protocols, which sign only one bit at a time and consume resources in a linear fashion with the document volume increasing, the signature rate of our OTHU protocol has a great enhancement. Moreover, the success probability of attacks from the external increases linearly as the document volume increases, which is discussed in detail in Appendix \ref{Appendix3}. 
Given the OTUH method, our protocol is unconditionally secure, allowing the imperfection of the secret keys distributed. This removes the necessity for privacy amplification. }

Our approach ensures that the shared keys we utilize are immune to detector side-channel attacks. This is accomplished by the incorporation of the MDI concept~\cite{32:Lo2012Measurement}. 
At the heart of our protocol lies the implementation of the asynchronous two-photon interference strategy, \textcolor{black}{which leads to a significant enhancement in the signature rates and an extension of the signature distance.} 
According to the OTUH, our protocol is robust to the document volume and we can attain signature rates that are several orders of magnitude higher without the need for perfect keys when the document volume is large, compared to the MDI signature schemes without OTUH~\cite{62:Puthoor2016Measurement}.
Furthermore, when compared to the twin-field scheme with single-photon interference referenced in Ref.~\cite{82:Li2023One-time}, our asynchronous MDI scheme holds an advantage as it does not require global phase tracking and phase locking. This implies that our protocol is not only easier to implement but also stands as a more practical scheme for future quantum networks. 
We analyze the formation process of shared keys, and we demonstrate the variations of $H_{\text{min}}^{\varepsilon}$ and $H_{\text{max}}^{\varepsilon_{\text{cor}}}$ with the signature distance by simulation. During this demonstration, we clearly reveal the formation process of these shared keys. This is based on the existing relationship between these quantum entropies and the unknown information to a potential attacker. 
By conducting simulations and comparisons, we have been able to demonstrate the significant performance of our approach, as well as clearly illustrating the formation process of the signatures utilized in our protocol. 

The structure of the article is as follows. In Sec. \ref{Sec2}, we introduce the content of our protocol, including the process of distribution and messaging. In Sec. \ref{Sec3}, we simulate and analyze the formation process of the shared keys during the distribution stage, and we demonstrate the composition of the raw key. Then we compare the performance of our asynchronous MDI-QDS protocol with the MDI-QDS described in Ref.~\cite{62:Puthoor2016Measurement} to emphasize the excellence of our protocol. In Sec. \ref{Sec4}, the article is concluded.

\begin{figure}
    \centering
    \includegraphics[page=1,width=\linewidth,keepaspectratio]{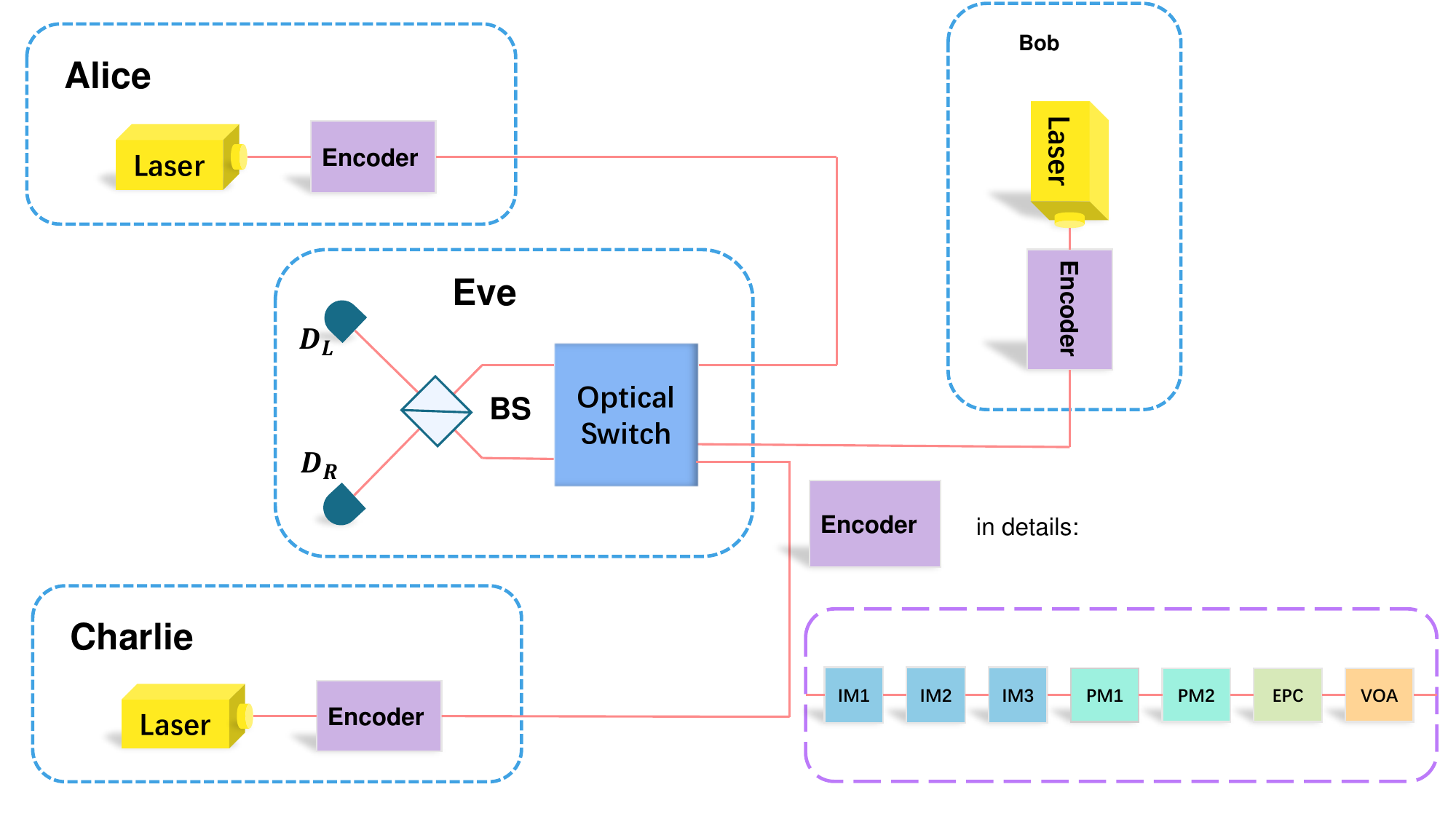}
    \caption{Schematic of the setup of the distribution stage of the proposed QDS protocol. Everyone generates weak coherent pulses with their own independent ultrastable lasers without mutual phase tracking. After encoding, they will send the pluses to Eve, who will perform the interference measurement and records successful clicks. The encoder consists of three intensity modulators, two phase modulators, an electrically driven polarization controller, and a variable optical attenuator. IM represents intensity modulator, PM represents phase modulator, EPC represents electrically driven polarization controller, and VOM represents variable optical attenuator. There is an optical switch in the node of Eve that can switch and select between different optical paths.}
    \label{f1}
\end{figure}
\section{PROTOCOL CONTENT}\label{Sec2}
\subsection{Distribution stage}
Our protocol employs the asynchronous MDI-KGP scheme for sharing keys among participants. In the distribution stage, we assume that in this three-party procedure, the matters of Alice-Bob and Alice-Charlie are independent and can be executed separately. \textcolor{black}{The setup is shown in Fig. \ref{f1}.}

\subsubsection{\textbf{Preparation}}
Consider each time slot $i\in{\{1,2,\ldots,N\}}$. Alice and Bob each prepare a weak laser pulse $\ket{e^{i\theta_{a(b)}}\sqrt{k_{a(b)}}}$ independently. Here, $\theta_{a(b)}$ is a phase value derived from $2\pi m_{a(b)}/M$, where $m_{a(b)}\in{\{0,1,\ldots,M-1\}}$, and $k_{a(b)}$ is an intensity chosen from the set $\{\mu_{a(b)},\nu_{a(b)},o_{a(b)}\}$ with the probabilities $p_{\mu_{a(b)}},p_{\nu_{a(b)}}$ and $p_{o_{a(b)}}=1-p_{\mu_{a(b)}}-p_{\nu_{a(b)}}$. 
The intensities within this set correspond to the signal, decoy, and vacuum state, in that order. 
Following this preparation phase, Alice and Bob transmit their pulses to a measurement node, referred to as Eve, via insecure channels. Although a similar process is also conducted between Alice and Charlie, we focus solely on the interaction between Alice and Bob in our discussion for simplicity.
\subsubsection{\textbf{Measurement and click filtering}}
For each bin, Eve conducts an interference measurement on the received pulses and logs the successful click events. Subsequently, she broadcasts the successful clicks along with the corresponding detector that registered the click. Following this, Alice and Bob publicly declare the events where they applied the decoy intensity $\nu_{a(b)}$ to the transmitted pulse. A click filtering process is then carried out, resulting in the discarding of clicks $(\mu_a|\nu_b)$ and $(\nu_a|\mu_b)$. All other clicks, apart from those discarded, are retained. 
\subsubsection{\textbf{Coincidence pairing}}
Our protocol does not pair pulses sent simultaneously as coincidences. Instead, we adopt a strategy that avoids the need for global phase tracking and phase locking. For the clicks we retain, we pair them with the nearest clicks within a time interval $T_c$ to form successful coincidences. If we fail to find a nearest click for a given click, we discard it. Upon successfully pairing coincidences, Alice and Bob calculate the total intensity$k^{\text{tot}}_{a(b)}$ of the two time bins they used. They also compute the phase difference between the earlier time bin ($e$) and the later time bin ($l$), denoted as $\phi_{a(b)}=\theta^{l}_{a(b)}-\theta^{e}_{a(b)}$. We denote the set of coincidences $[k^{\text{tot}}_{a}, k^{\text{tot}}_{b}]$ as $S_{[k^{\text{tot}}_{a}, k^{\text{tot}}_{b}]}$.
\subsubsection{\textbf{Sifting}}
After computing their results, Alice and Bob announce $k^{\text{tot}}_{a(b)}$ and $\phi_{a(b)}$. They discard any results where the total intensity satisfies $k^{\text{tot}}_{a(b)} \ge \mu_{a(b)}+\nu_{a(b)}$. 
\textcolor{black}{For the Z-basis, Alice (Bob) extracts a bit 0 (1) if she (he) sends $\mu_{a(b)}$ in the early time bin and $o_{a(b)}$ in the late time bin. Otherwise, Alice (Bob) extracts an opposite bit.}

For the X-basis, we use coincidences $[2\nu_a,2\nu_b]$ to extract bits. Alice and Bob first calculate $\phi_{ab}=\phi_a-\phi_b$, which represents the phase difference between the phase difference of Alice and Bob in the early time and the later time. They then calculate $\phi=\phi_{ab}\mod2\pi$. \textcolor{black}{If the result is 0 or $\pi$, Alice and Bob will extract 0 in the X-basis. If the result is 0 and both detectors click, Bob will flip the bit. If the result is $\pi$ and only a detector clicks and the same detector clicks twice, Bob will flip too.} If the result is other values except 0 and $\pi$, we will discard this coincidence. 
\subsubsection{\textbf{Parameter estimation}}
Alice and Bob can then obtain their own raw key from the Z-basis, which has the length of $n_z$. The parameters $s^z_0,s^z_{11},\phi^z_{11}$ will also be computed and retained. 
These parameters represent the length of the bits derived from the vacuum events, single-photon events, and the phase-error rate of the single-photon events, respectively. The error rate of the bits in the Z-basis $E_z$ will also be computed. 
The details of the estimation could be found in Appendix \ref{Appendix2}. All these are useful in post-processing, which will help to get the length of shared keys and the signature. 

\begin{figure*}[ht]
    \centering
    \includegraphics[page=1,width=\textwidth,keepaspectratio]{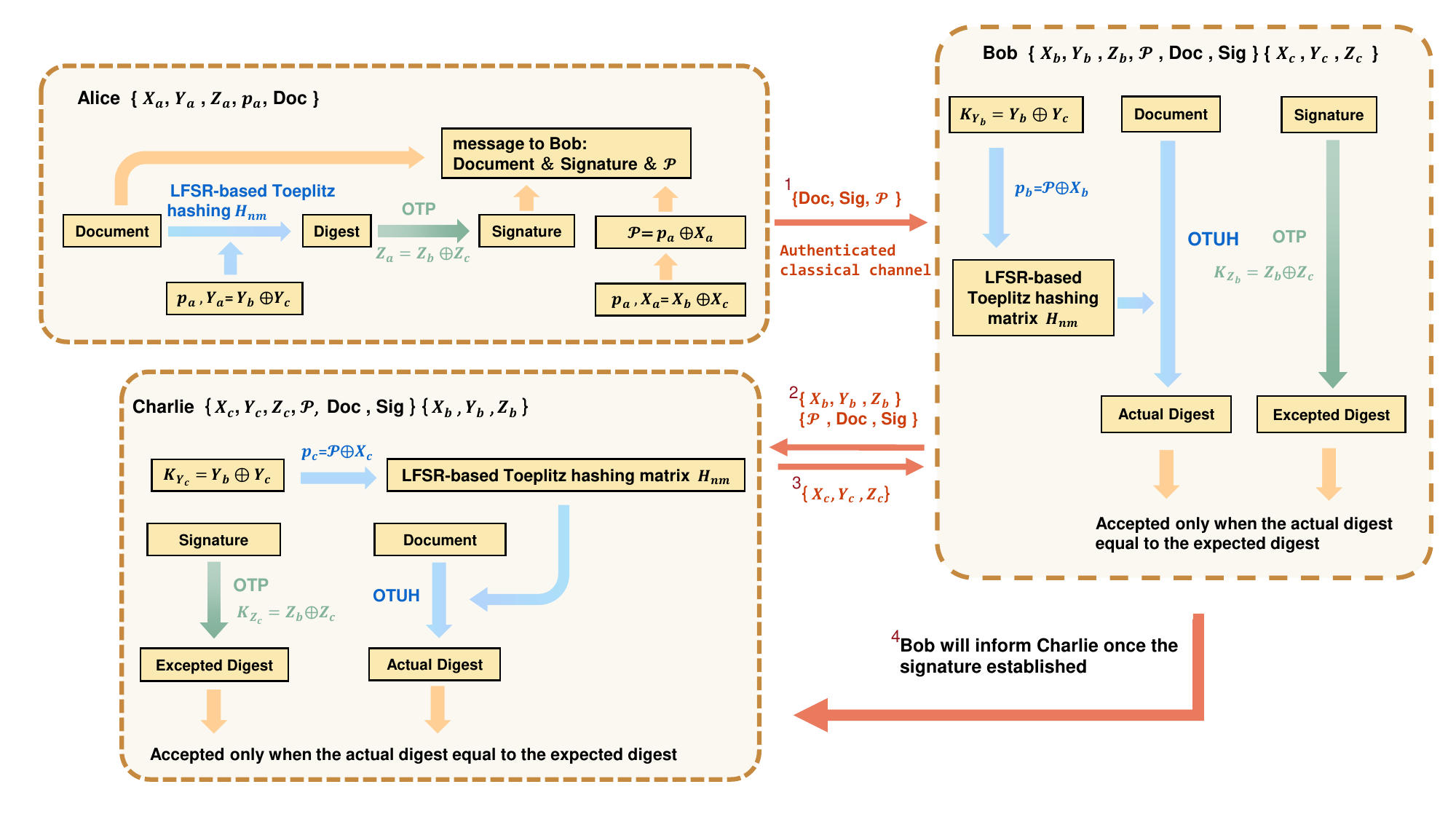}
    \caption{Schematic of the implementation of the messaging stage of the proposed QDS protocol. It is carried out between three participants, which communicate with each other through authenticated classical channels. Firstly, Alice uses the strings $\{Y_a, p_a\}$ to generate an LFSR-based Toeplitz hashing matrix $H_{nm}$, and then uses the hashing function to encrypt the document, getting the digest. Then she uses $Z_a$ and the digest to obtain the signature through \textcolor{black}{one-time pad (OTP),} and she encrypts $p_a$ with $X_a$, getting $p$. After this she sends $\{\text{Doc},\text{Sig},p\}$ to Bob. On realizing the information from Alice, Bob will communicate with Charlie and he will use the LFSR-based Toeplitz hashing matrix $H_{nm}$ generated from $K_{X_b}$ and $p_b$ to encrypt the document to get the actual digest. Meanwhile, he uses $K_{Z_b}$ and the signature to get the excepted digest. Comparing the two digests, he will decide whether to accept the signature and inform the result to Charlie. If Bob accept the signature, he will inform the result to Charlie. Then, Charlie will perform a similar verification process to that of Bob, to verify the validity of the signature.}
    \label{f2}
\end{figure*}

\subsubsection{\textbf{Error correction}}
After obtaining the raw key, Alice and Bob will distill it using error correction with a correction factor of $\varepsilon_{\text{cor}}$~\cite{95:Brassard1994Secret-key,96:Yan2008Information}. 
The length of keys will remain $n_z$, and the unknown information to a potential attacker will be represented as $\mathcal{H}$~\cite{82:Li2023One-time}. During this stage, there is no need to perform privacy amplification. Subsequently, Alice randomly disturbs the order of the key and announces the new order to Bob through an authenticated channel. This will allow them both to obtain the final key. These keys will then be divided into several strings of $n$-bits, which will play \textcolor{black}{an} important role in the messaging stage. 
\newline

The entire distribution process will also involve both Alice and Charlie. For the sake of simplicity, we did not previously mention that the keys of a certain length are also distributed between them. Once these keys have been distributed, they will be divided into several segments, each of which will be used for specific operations in the subsequent process. 

\subsection{Messaging stage}
In this section, we demonstrate the key aspect of the protocol, which is to perfectly correlate the bits among three parties, as described in Ref.~\cite{82:Li2023One-time}. This requires an asymmetric key relationship among the three parties. We use one-time almost XOR universal$_2$ (AXU) hashing, specifically, \textcolor{black}{the Linear Feedback Shift Register (LFSR)}-based Toeplitz hashing, to generate the protocol’s signature. 
The strings of length $n_z$ on the sides of Alice, Bob, and Charlie have already been divided into segments of length $n$. These segments are denoted as $\{X_a,X_b,X_c,Y_a,Y_b,Y_c,Z_a,Z_b,Z_c\}$, each of which has a length of $n$. The subscripts $\{a,b,c\}$ indicate that the string belongs to Alice, Bob, or Charlie, respectively. These strings satisfy the equations 
\[X_a=X_b\oplus X_c,\]
\[Y_a=Y_b\oplus Y_c,\]
\[Z_a=Z_b\oplus Z_c.\] 
We will use these strings to execute the protocol between the three parties. \textcolor{black}{And the schematic of the messaging stage is shown in Fig. \ref{f2}.}
\subsubsection{\textbf{Signing of Alice}}
Alice holds a set of $n$-bit long strings $\{X_a,Y_a,Z_a\}$. 
First, she uses a quantum random number generator to produce an $n$-bit long random string, which is called $p_a$. This string is used to create a monic irreducible polynomial $p(x)$ of order $n$ in GF$(2)$. 
Second, Alice uses the bit string $Y_a$ and the irreducible polynomial (quantum random number $p_a$) to generate a random linear feedback shift register-based (LFSR-based) Toeplitz matrix $H_{nm}$, which has $n$ rows and $m$ columns. She applies this matrix to the $m$-bit document Doc, resulting in an $n$-bit hash value $\text{Dig}=H_{nm}\cdot \text{Doc}$. 
Third, Alice encrypts Dig using $Z_a$ to obtain the final signature $\text{Sig}=\text{Dig}\oplus Z_a$. 
In addition, Alice encrypts $p_a$ by $X_a$ to get $\mathcal{P}=p_a\oplus X_a$. Fourth, Alice transmits the set $\{\text{Sig},\mathcal{P},\text{Doc}\}$ to Bob through an authenticated classical channel. 
\subsubsection{\textbf{Verification of Bob}}
Upon receiving the signal from Alice, Bob transmits $\{\text{Sig},\mathcal{P},\text{Doc}\}$ and $\{X_b,Y_b,Z_b\}$ to Charlie. After receiving the signal from Bob, Charlie transfers $\{X_c,Y_c,Z_c\}$ to Bob. At this point, Bob has the set of strings $\{\text{Sig},\mathcal{P},\text{Doc},X_b,Y_b,Z_b,X_c,Y_c,Z_c\}$, which will be used to perform the verification stage. All data are transmitted through an authenticated channel. 
First, Bob generates the new strings $\{K_{X_b}=X_b\oplus X_c,K_{Y_b}=Y_b\oplus Y_c,K_{Z_b}=Z_b\oplus Z_c\}$ via XOR operation. 
Second, using $K_{X_b}$ and $K_{Z_b}$, Bob obtains $p_b$ and the expected digest via XOR decryption. Then, with $K_{Y_b}$, Bob uses it and $p_b$ to form an LFSR-based Toeplitz matrix, and obtains the actual digest via a hash operation with the matrix. 
Third, Bob accepts the signature if the actual digest equals the expected digest, and then informs Charlie of this result. If the two digests are not identical, he will reject the signature and announces the protocol’s abortion. 
The signature will be established if Bob accepts it, and the establishment of the signature does not require consideration of Charlie, who plays the role of a notary. 
\subsubsection{\textbf{Verification of Charlie}}
If Charlie receives a successful signal from Bob, he will perform the verification stage just like Bob. At this point, Charlie has the same set of strings as Bob, which is $\{\text{Sig},\mathcal{P},\text{Doc},X_b,Y_b,Z_b,X_c,Y_c,Z_c \}$. 
First, Charlie generates the new strings $\{K_{X_c}=X_b\oplus X_c,K_{Y_c}=Y_b\oplus Y_c,K_{Z_c}=Z_b\oplus Z_c\}$ via the XOR operation. 
Second, He exploits $K_{X_c}$ and $K_{Z_c}$ to obtain the expected digest and string $p_c$ via XOR decryption. Then, Using $K_{Y_c}$, he obtains the actual digest via a hash operation like Bob. 
Third, if the two digests are identical, he will accept the protocol; otherwise, he will reject it.
\newline

Under this framework, various AXU hash functions could be employed to play a major role. In our protocol, we specifically exploit the LFSR-based Toeplitz hashing, which is a fantastic function that can map a document of any length to a fixed length.

From the description above, we know that in order to sign a message of m-bits length, Alice should distribute six bit strings $X_b,Y_b,Z_c$ to Bob, and $X_a,Y_c,Z_c$ to Charlie. The subscript indicates the participant performing the KGP with Alice, where $b$ represents Bob and $c$ represents Charlie. 
We set the fixed length of strings as $n$. With each channel generating three strings, and the length $n_Z$ of the raw key distributed in each channel, we could calculate the signature rate~\cite{82:Li2023One-time}:
\begin{equation}
R_{\text{sig}}=\frac{n_z}{3n}.
\label{Eq1}
\end{equation}

\section{SIMULATION AND DISCUSSION}\label{Sec3}
\begin{table}[b]
    \centering
    \begin{tabular}{ccccccc}
    \hline
    \hline
    $\eta_d$   & $p_d$               & $f$ & $\alpha_f$ & $e_d$  & $\varepsilon$     & $F$       \\
    \hline
    $80\%$      & $2.5\times10^{-10}$ & 1.1 & 0.16      & 0.04   & $1\times10^{-10}$ & $1\text{GHz}$    \\
    \hline
    \hline
    \end{tabular}
\caption{This table contains the parameters of the simulation we set, in which $\eta_d$ and $p_d$ 
represents the detection efficiency and the dark count rate of the detectors we use. $f$ is the 
error correction efficiency. $e_d$ represents the misalignment error rate, and $\alpha_f$ is the
attenuation coefficient of the fiber. The parameter $\varepsilon$ is the value of the variables 
$\varepsilon^{\prime}$, $\hat{\varepsilon}$ and $\varepsilon_{\text{cor}}$. $F$ is the system clock frequency}
\label{t1}
\end{table}

During the distribution stage, we have performed the parameter estimation and error correction. After the distribution stage, the unknown information to a possible attacker $\mathcal{H}$ could be expressed with the smooth min-entropy and the smooth max-entropy as:
\begin{equation}
    \begin{split}  
        \mathcal{H} \ge H_{\text{min}}^{\varepsilon}-H_{\text{max}}^{\varepsilon_{\text{cor}}},
    \end{split}
    \label{Eq2}
\end{equation}
in which the $H_{\text{min}}^{\varepsilon}$ and the $H_{\text{max}}^{\varepsilon_{\text{cor}}}$ could be separately 
expressed as 
\begin{equation}
    \begin{split}  
        H_{\text{min}}^{\varepsilon} \ge s_{0}^z+s_{11}^z[1-H(\phi_{11}^z)]
        -2\text{log}_2(\frac{2}{\varepsilon^{\prime}\hat{\varepsilon}}),
    \end{split}
    \label{Eq3}
\end{equation}
\begin{equation}
    \begin{split}  
        H_{\text{max}}^{\varepsilon_{\text{cor}}}=n_zfH(E_z)+\text{log}_2(\frac{2}{\varepsilon_{\text{cor}}}),
    \end{split}
    \label{Eq4}
\end{equation}

\begin{figure}[t]
    \includegraphics[width=\linewidth]{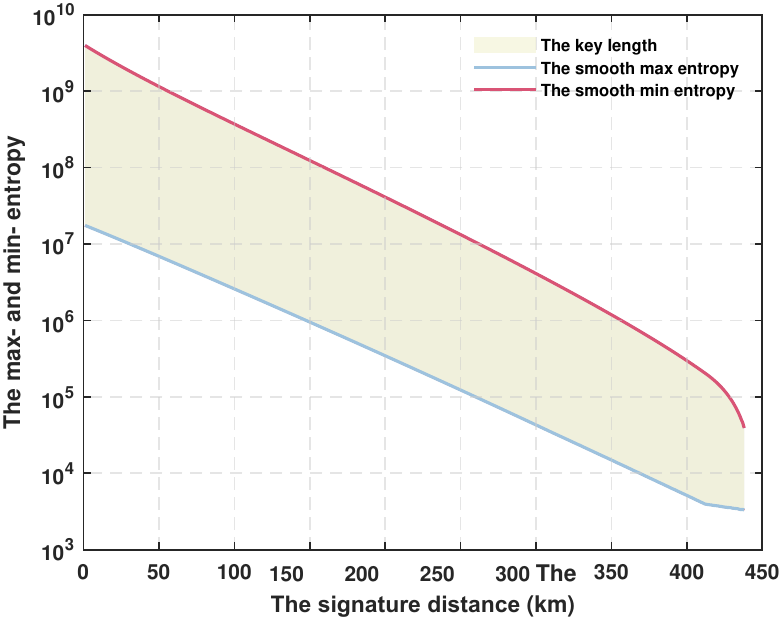}
    \caption{The schematic of the variation of the smooth entropies $H_{\text{min}}^{\varepsilon}$ and $H_{\text{max}}^{\varepsilon_{\text{cor}}}$ with distance $l$ and \textcolor{black}{the colored area as the legend} represents $\mathcal{H}$, the portion ultimately unknown to a possible attacker. Obtained by simulating the distribution stage with the parameters in Table \ref{t1}.}
 \label{f3}
 \end{figure}

where $f$ is the error correction efficiency, $s_{0}^z$ is the number of vacuum events, $s_{11}^z$ is the number of single-photon pairs event, $\phi^{z}_{11}$ represents the number of the phase error rate of single-photon pairs, and $E_z$ is the bit error rate of Z-basis during the distribution stage. The function is the  binary Shannon entropy function, which could be expressed as: 
\begin{equation}
    H(x)=-x\text{log}_2x-(1-x)\text{log}_2(1-x),
\end{equation}
Using these two entropies, we could get the length of $\mathcal{H}$:
\begin{equation}
\begin{split}  
    \mathcal{H}&\ge s_{0}^z+s_{11}^z[1-H(\phi_{11}^z)]-n_zfH(E_z)\\
    &\quad -2\text{log}_2(\frac{2}{\varepsilon^{\prime}\hat{\varepsilon}})-\text{log}_2(\frac{2}{\varepsilon_{\text{cor}}}),
\end{split}
\label{Eq6}
\end{equation}
of which the details will be introduced in Appendix \ref{Appendix4}, which involves the details of these smooth entropies. 

In order to delve deeper into the dimensionality of $\mathcal{H}$, we separately examined the two key components, $H_{\text{min}}^{\varepsilon}$ and $H_{\text{max}}^{\varepsilon_{\text{cor}}}$. This included an analysis of the variations in their numerical values and the changes in the percentage they represent in the raw key $n_z$. 
In this context, we set the $N$ to $10^{12}$, which represents the total number of transmitted pulse pairs. The parameters of the simulation we set could be found in Table \ref{t1}. 

\begin{figure}[t]
    \includegraphics[width=\linewidth]{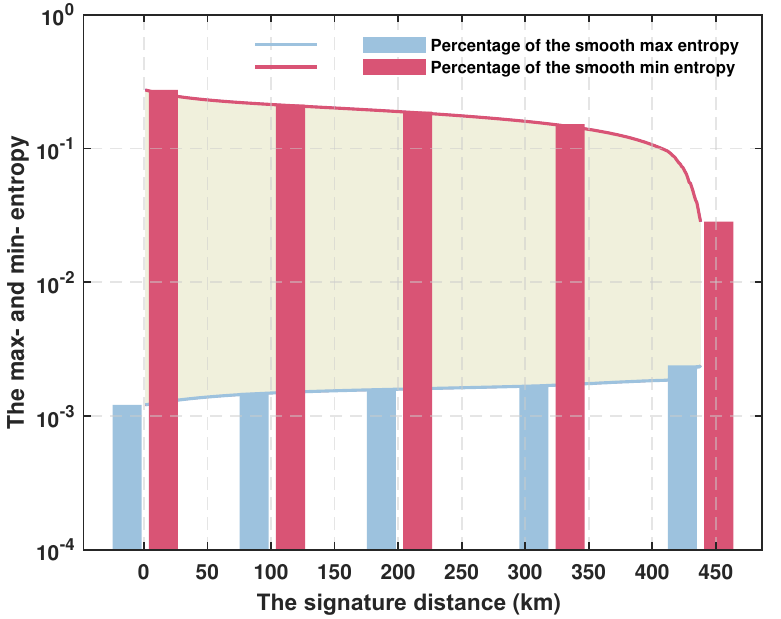}
    \caption{The schematic of the variation of percentage of the smooth entropies $H_{\text{min}}^{\varepsilon}$ and $H_{\text{max}}^{\varepsilon_{\text{cor}}}$ with distance $l$. Obtained by simulating the distribution stage with the parameters in Table \ref{t1}. \textcolor{black}{The colored area as the legend} refers to the percentage of $\mathcal{H}$.}
\label{f4}
 \end{figure}

By simulating the implementation of the distribution stage with these parameters, we are able to observe the variation of the absolute values of the smooth min- and max-entropies, $H_{\text{min}}^{\varepsilon}$ and 
$H_{\text{max}}^{\varepsilon_{\text{cor}}}$ with respect to distance $l$. $H_{\text{min}}^{\varepsilon}$ represents the maximum length of a bit string that can be computed from the raw key before error correction, which is $\varepsilon$-closing to a perfectly uniform string. 
This string is independent of the side information eavesdropped by Eve. $H_{\text{max}}^{\varepsilon_{\text{cor}}}$ represents the amount of information consumed in error correction.

The shaded area between two curves represents the unknown information $\mathcal{H}$. As can be seen in Fig. \ref{f3}, with the increase in 
the distance, the absolute value of $H_{\text{min}}^{\varepsilon}$ and $H_{\text{max}}^{\varepsilon_{\text{cor}}}$ decreased by a similar slope. However, since this is a semi-logarithmic plot with the y-axis on a logarithmic scale, $\mathcal{H}$ was decreasing exponentially. Towards the end of Fig. \ref{f3}, the length of $\mathcal{H}$ experienced a sharp decrease, corresponding to the drop-off of the rates of the KGP process. As the attenuation of signals increases to a significant degree, the total amount of information that can be transmitted decreases substantially. Concurrently, the influence of noise becomes increasingly significant. This results in the observed drop-off. This process could be seen more intuitively in Fig. \ref{f4}.

In Fig. \ref{f4}, we illustrate the variation of the percentage of the smooth min-entropy $H_{\text{min}}^{\varepsilon}$ and the smooth max-entropy $H_{\text{max}}^{\varepsilon_{\text{cor}}}$ occupied in the raw key with respect to distance $l$. The percentage of the smooth min-entropy $H_{\text{min}}^{\varepsilon}$ shows a slight decrease, but overall, it remains almost unchanged before 410 km, and the percentage of the smooth max-entropy $H_{\text{max}}^{\varepsilon_{\text{cor}}}$ shows a very slight increase, with almost no change before 410 km as well. 
After 410 km the percentage of $H_{\text{min}}^{\varepsilon}$ undergoes a sharp decrease. This is primarily due to the reduced number of pulses that reach this distance, coupled with the increasingly pronounced impact of noise. The combined sum of these two entropies was notably less than 1. This is attributed to the constant need to discard a certain amount of information before error correction, specifically $(1-H_{\text{min}}^{\varepsilon})$, to maintain security against potential external threats.

Given the relationship between entropy and information~\cite{92:Konig2009Operational}, we apply this principle within quantum systems as well as hybrid classical-quantum systems to generate keys and estimate signature length, thereby ensuring security. This is precisely where QDS protocols distinguish themselves from classical ones, as well as in the characteristic of not requiring assumptions of an authenticated broadcast channel or a trusted authority~\cite{9:Amiri2015Unconditionally,86:Chaum1991Unconditionally,87:Hanaoka2000UnconditionallySD}.

\begin{figure}[t]
    \includegraphics[width=\linewidth]{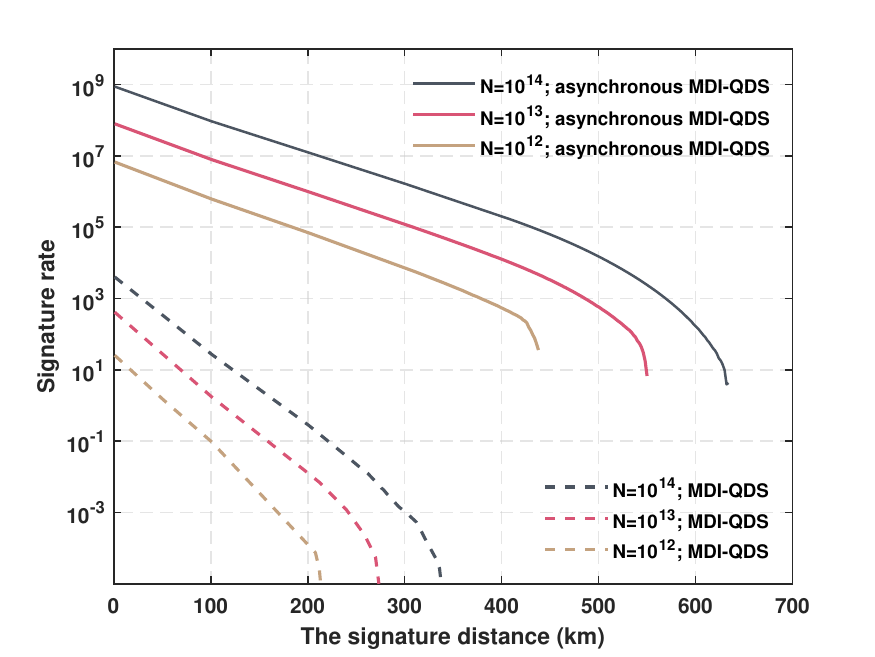}
    \caption{Comparison of signature rates of our proposed  asynchronous MDI-QDS protocol and the MDI-QDS described in Ref.~\cite{62:Puthoor2016Measurement} under different data size $N$ of $10^{12}$, $10^{13}$, and $10^{14}$. The message is assumed to be $10^3$ bits. Security bound of the signatures is $10^{-10}$. Other parameters of this simulation are consistent with those in Table \ref{t1}. }
\label{f5}
\end{figure}

To showcase the superior performance of our protocol, we conducted simulations comparing our protocol with the MDI-QDS~\cite{62:Puthoor2016Measurement}. 
For reasonable comparison, we use the best known MDI-KGP method to distribute quantum states used for MDI-QDS~\cite{62:Puthoor2016Measurement}, i.e., a four-intensity decoy-state protocol with the double-scanning method~\cite{88:Jiang2021Higher}.
These simulations were performed under varying data sizes $N$ of $10^{12}$, $10^{13}$, 
and $10^{14}$, with the document message size capped at $10^3$ bits. The results of this simulation could be seen in Fig. \ref{f5}.

In this simulation, it is demonstrated that the maximum signature distance of the proposed protocol is extended by approximately two times compared to MDI-QDS. The substantial improvement observed can be attributed to the implementation of the asynchronous two-photon interference strategy. 
During the distribution stage, we asynchronously pair two successful clicks within a long pairing time. These asynchronous pairs are then used to generate the key for messaging and signature. This approach aids in breaking through the secret-key capacity barrier without the need for global phase locking~\cite{58:Zhou2023Experimental} during the distribution stage. As a result, the distribution distance of the distribution stage is approximately doubled compared to the MDI-QDS. 

As depicted in Fig. \ref{f5}, when compared to the MDI-QDS~\cite{62:Puthoor2016Measurement} that does not incorporate OTUH, the signature rates of our proposed  asynchronous MDI-QDS protocol are enhanced by six to seven orders of magnitude. The observed enhancement is derived from the advantages of our OTUH scheme, of which the details concerning the secure information in the raw key, denoted as $\mathcal{H}$, have been thoroughly discussed in Sec. \ref{Sec3}. 
Our OTUH scheme is capable of projecting a document containing a large volume of information to an adjustable hash value. Consequently, our protocol is not sensitive to the size of the document and can perform more effectively when handling documents of larger sizes.

\section{CONCLUSION}\label{Sec4}
On the whole, we propose an asynchronous MDI-QDS protocol with OTUH, which could \textcolor{black}{achieve a higher signature rate} and longer signature distance than other schemes. 
In our paper, we delve into the composition of the raw key and explore the relationship between its various components, entropy, and information. This analysis provides a comprehensive understanding of the formation process of the shared keys in our QDS protocol and offers profound insights into the OTUH-QDS process. 
By simulating and comparing our proposed protocol with the MDI-QDS described in~\cite{62:Puthoor2016Measurement}, it turns out that our protocol has significant improvements in terms of signature rates and distance due to the applications of OTUH and the asynchronous two-photon interference strategy. 
By employing the asynchronous two-photon interference strategy~\cite{55:Xie2022Breaking}, the maximum signature distance can be significantly extended, potentially up to twice the distance without the asynchronous two-photon interference strategy, because of the reduced channel loss. With OTUH employed, our protocol has strong robustness against the document volume. This makes our protocol have a significant performance when handling extensive documents, especially several orders of magnitude higher compared to the MDI-QDS without OTUH.
Furthermore, our protocol does not need global phase tracking and phase locking compared to the twin-field scheme with single-photon interference referenced in Ref.~\cite{82:Li2023One-time}, thus making our protocol more practical and easier to implement. The feasibility of the asynchronous distribution scheme has been experimentally qualified ~\cite{58:Zhou2023Experimental}, which means that the realization of our proposed protocol is easier and not far from reality. 

\section*{ACKNOWLEDGMENTS}
This study was supported by the National Natural Science Foundation of China (No. 12274223) and the Program for Innovative Talents and 
Entrepreneurs in Jiangsu (No. JSSCRC2021484).

\appendix
\section{LFSR-BASED TOEPLITZ HASH FUNCTION}\label{Appendix1}
An $(m,n)$-family $H$ of hash functions is a collection of functions that map the set of binary strings of 
length $m$ into the set of binary strings of length $n$~\cite{89:Krawczyk1994LFSR-based}. The LFSR-based Toeplitz hash function can be expressed as 
\begin{equation}
h_{p,s}(M)=H_{nm}\cdot M,
\end{equation}
which can map the binary string $M$ of length $m$ to a binary string $h_{p,s}(M)$ of length $n$, and the LFSR-based Toeplitz 
matrix $H_{nm}$ is a matrix of size n by m constructed from an irreducible polynomial $p(x)$ over GF(2) of degree $n$
and an initial state $s$.

The $m$-bits message $M$ can be represented as $(M_0, M_1, \cdots, M_{m-1})^T$; the initial state $s$ can be denoted 
as $(S_n, S_{n-1}, \cdots, S_1)^T$, and $p(x)$ is an irreducible polynomial over GF(2) of degree n, which can be 
expressed as $p(x)=x^n+p_{n-1}x^{n-1}+\cdots+p_1x+p_0$. This polynomial is obviously characterized by its coefficients 
of the order of $x$ from $0$ to $n-1$, so we could rewritten it as $p=(p_{n-1}, p_{n-2}, \cdots, p_1, p_0)^T$. The matrix 
$H_{nm}$ could be constructed from $s$ and $p$ as follows~\cite{89:Krawczyk1994LFSR-based,82:Li2023One-time}:

First, we need to define an n-by-n matrix $W$ which is solely determined by the $p$.
\begin{equation}
    W=\begin{pmatrix}
 p_{n-1} & p_{n-2} & \cdots & p_1    & p_0    \\
 1       & 0       & \cdots & 0      & 0      \\
 0       & 1       & \cdots & 0      & 0      \\
 \vdots  & \vdots  & \ddots & \vdots & \vdots \\
 0       & 0       & \cdots & 1      & 0
    \end{pmatrix},
\end{equation}

From the definition of the matrix $W$, we could find that, $p(x)$ is the characteristic polynomial of the matrix $W$. 
Then according to Hamilton-Cayley theorem, $p(W)=0$~\cite{90:Mertzios1986Cayley-Hamilton}.

Applying this matrix $W$ to the vector $s$, we could get $s_1=(S_{n+1}, S_n, \cdots, S_2)^T$, where $S_{n+1}=p\cdot s$. 
We could see that the function of the matrix $W$ is to shift down each element of the vector s and prepend a new element $p\cdot s$. 

Repeating this operation $m-1$ times and denoting the vector $s$ as $s_0$, we can get a set of vectors 
$\{s_0, s_1, \cdots, s_{m-1}\}$ satisfying:
\begin{equation}
    s_{i+1}=W\cdot s_i,
\end{equation}

Since $s_0=s$, we could express each element of this set with $W$ and $s$ as:
\begin{equation}
    s_i=W^i\cdot s \ \ \ (0\le i\le m-1),
\end{equation}

So we could get an n-by-m matrix $(s_0, s_1, \cdots, s_{m-1})$, which has the ability to map 
a $m$-bits vector to a $n$-bits vector. This matrix is the LFSR-based Toeplitz 
matrix $H_{nm}$ we want.
\begin{equation}
H_{nm}=(s_0, s_1, \cdots, s_{m-1}),
\end{equation}

We  we can rewrite the function as:
\begin{equation}
    \begin{split}
    h_{p,s}(M)&=H_{nm}\cdot M \\
              &=(s_0, s_1, \cdots, s_{m-1}) \cdot \begin{pmatrix}
                                                 M_0    \\
                                                 M_1    \\
                                                 \vdots \\
                                                 M_{m-1}
                                                \end{pmatrix}\\
              &=\mathcal{M_W}(W)\cdot s,
    \end{split}
\end{equation}

in which we have:
\begin{equation}
    \begin{split}
        \mathcal{M_W}(M)&=M_{m-1}\cdot W^{m-1}+M_{m-2}\cdot W^{m-2}+\cdots \\
                      &+M_1\cdot W+M_0,
   \end{split}
\end{equation}

So, if $p(x)|\mathcal{M_W}(x)$, $\mathcal{M_W}(W)$ will be equal to $0$, and then $h_{p,s}(M)=0$.
\section{CALCULATION OF PARAMETERS}\label{Appendix2}
According to Eq. \eqref{Eq1}, to calculate the signature rate $R_{\text{sig}}$, we need to calculate the length of raw key $n_z$ and the length of the signature $n$ after the distribution stage.

For the purpose of calculating these two parameters, there exist some parameters we need to estimate during the distribution stage, which includes the lower bound of vacuum events and single-photon pairs in the Z basis $\underline{s}_0^z$ and $\underline{s}_{11}^z$; the upper bound of the phase error rate $\overline{\phi}_{11}^z$; the length of the raw key $n_z$; and the bit error rate in the Z basis $E_z$. 

The overline and the underline represent the Chernoff bounds of the variables, which could be introduced as below~\cite{58:Zhou2023Experimental,91:yin2020tight}:

Let $x$ represent the observed value and $x^{*}$ represent the expected value, and we have the upper and lower 
bounds of the observed value~\cite{58:Zhou2023Experimental,91:yin2020tight}:
\begin{equation}
    \begin{split}
    \overline{x}&=O^U(x^*) \\
                &=x^{*}+\frac{\beta}{2}+\sqrt{2\beta x^{*}+\frac{\beta^2}{4}},
    \end{split}
\end{equation}
and
\begin{equation}
    \begin{split}
    \underline{x}&=O^L(x^*) \\
                 &=x^{*}-\sqrt{2\beta x^{*}},
    \label{B2}
    \end{split}
\end{equation}
and the upper and lower bounds of the expected value:
\begin{equation}
    \overline{x}^{*}=x+\beta+\sqrt{2\beta x+\beta^2},
\end{equation}
and
\begin{equation}
    \underline{x}^{*}=\max\{x-\frac{\beta}{2}-\sqrt{2\beta x+\frac{\beta^2}{4}},\ 0\},
\end{equation}
where $\beta=\ln \epsilon^{-1}$.

Furthermore, the random sampling theorem will also be applied in our calculation, which is given as below~\cite{58:Zhou2023Experimental,91:yin2020tight}: 
\begin{equation}
    \overline{\chi}\leq \lambda+ \gamma^U( n, k, \lambda, \epsilon),
\end{equation}
where 
\begin{equation}
    \gamma^U( n, k, \lambda, \epsilon)=\frac{\frac{(1-2\lambda)AG}{n+k}+\sqrt{\frac{A^2G^2}{(n+k)^2}+4\lambda(1-\lambda)G}}{2+2\frac{A^2G}{(n+k)^2}},
\end{equation}
in which 
\begin{equation}
    A=\max\{n,k\},
\end{equation}
and
\begin{equation}
    G=\frac{n+k}{nk}\ln(\frac{n+k}{2\pi nk\lambda(1-\lambda)\epsilon^2}).
\end{equation}

When Alice and Bod send When Alice and Bob send intensities $k_a$ and $k_b$ with phase difference $\theta$, the gain corresponding to only detector $L$ and $R$ click can be represented as below~\cite{58:Zhou2023Experimental}: 
\begin{equation}
    \begin{split}
    q_{(k_a|k_b)}^{\theta, L}&=y_{(k_a|k_b)}^R e^{\eta_d^R \sqrt{\eta_a k_a \eta_b k_b} \cos\theta} \cdot \\
                             & \times (1-y_{(k_a|k_b)}^L e^{-\eta_d^L \sqrt{\eta_a k_a \eta_b k_b}} \cos\theta),
    \end{split}
\end{equation}

\begin{equation}
    \begin{split}
    q_{(k_a|k_b)}^{\theta, R}&=y_{(k_a|k_b)}^L e^{-\eta_d^L \sqrt{\eta_a k_a \eta_b k_b} \cos\theta} \cdot \\
                             & \times (1-y_{(k_a|k_b)}^R e^{\eta_d^R \sqrt{\eta_a k_a \eta_b k_b}} \cos\theta),
    \end{split}
\end{equation}
in which $\eta_{a(b)}=10^{-\frac{\alpha l_{a(b)}}{10}}$, and
\begin{equation}
    \begin{split}
        y_{(k_a|k_b)}^{L(R)}=(1-p_d^{L(R)})\cdot e^{-\frac{\eta_d^{L(R)}(\eta_a k_a+\eta_b k_b)}{2}},
    \end{split}
\end{equation}
where $\eta_d^{L(R)}$ and $p_d^{L(R)}$ represents the detection efficiency and the dark count rate of the detector $D_{L(R)}$ 
respectively. The overall gain $q_{(k_a|k_b)}$ can be expressed as:
\begin{widetext}
\begin{equation}
    \begin{split}
    q_{(k_a|k_b)}&=\frac{1}{2\pi}\int_0^{2\pi}(q_{(k_a|k_b)}^{\theta, L}+q_{(k_a|k_b)}^{\theta, R})d\theta \\
                 &=y_{(k_a|k_b)}^L I_0 (\eta_d^L \sqrt{\eta_a k_a \eta_b k_b}) 
                 +y_{(k_a|k_b)}^R I_0 (\eta_d^R \sqrt{\eta_a k_a \eta_b k_b}) 
                 -2y_{(k_a|k_b)}^L y_{(k_a|k_b)}^R \cdot I_0 [(\eta_L-\eta_R) \sqrt{\eta_a k_a \eta_b k_b}],
    \end{split}
\end{equation}
\end{widetext}
where $I_0(x)$ refers to the zero-order modified Bessel function of the first kind.

Denote the probability of having a click event as $q_{\text{tot}}$. Click filtering applied, $q_{\text{tot}}$ could be expressed as: 
\begin{equation}
    \begin{split}
    q_{\text{tot}}&=\sum_{k_a, k_b}{p_{k_a}p_{k_b}q_{(k_a|k_b)}}-p_{\mu_a}p_{\nu_b}q_{(\mu_a|\nu _b)} \\
           &-p_{\nu_a}p_{\mu_b}q_{(\nu_a|\mu _b)}.
    \end{split}
\end{equation}

The probability of at least one click event occurring following a given time bin with a click event within the time interval $T_c$ could be expressed as~\cite{58:Zhou2023Experimental}: 
\begin{equation}
    q_{T_c}=1-(1-q_{\text{tot}})^{N_{T_c}},
\end{equation}
where $N_{T_c}=FT_c$ is the number of time bins within the time interval $T_c$, and $F$ is the system clock frequency, which 
can be found in Table \ref{t1}.
Therefore, the total number of valid successful pairing results and the average of the pairing interval could be obtained: 
\begin{equation}
    n_{\text{tot}}=\frac{N q_{\text{tot}}}{1+1/q_{T_c}},
\end{equation}
\begin{equation}
    T_{\text{mean}}=\frac{1-N_{T_c} q_{\text{tot}}(1/q_{T_c}-1)}{F q_{\text{tot}}}.
\end{equation}

Having calculated these parameters above, $n_{[k_a^{\text{tot}}, k_a^{\text{tot}}]}$, the total number of set $S_{[k_a^{\text{tot}}, k_a^{\text{tot}}]}$, could be obtained~\cite{58:Zhou2023Experimental}. But this formula is inapplicable to the set $S_{[2\nu_a, 2\nu_b]}$. The total number of set $S_{[k_a^{\text{tot}}, k_a^{\text{tot}}]}(k_{a(b)}^{\text{tot}} \neq 2\nu_{a(b)})$ and $S_{[2\nu_a, 2\nu_b]}$ could be expressed respectively as follows:
\begin{widetext}
\begin{equation}
    \begin{split}
    n_{[k_a^{\text{tot}}, k_a^{\text{tot}}]}=n_{\text{tot}}  
                               \times \sum_{k_a^e+k_a^l=k_a^{\text{tot}}} \sum_{k_b^e+k_b^l=k_b^{\text{tot}}} 
                              {\frac{p_{k_a^e} p_{k_b^e} q_{(k_a^e|k_b^e)}}{q_{\text{tot}}} \frac{p_{k_a^l} p_{k_b^l} q_{(k_a^l|k_b^l)}}{q_{\text{tot}}}}.
    \end{split}
\end{equation}
\begin{equation}
    \begin{split}
    n_{[2\nu_a, 2\nu_b]}=\frac{n_{\text{tot}}}{M\pi}\cdot 
                    \int_0^{2\pi}{(\frac{p_{\nu_a}p_{\nu_b}q_{(\nu_a|\nu_b)}^\theta}{q_{\text{tot}}}\frac{p_{\nu_a}p_{\nu_b}q_{(\nu_a|\nu_b)}^\theta}{q_{\text{tot}}})} d\theta,
    \end{split}
\end{equation}
\end{widetext}
Furthermore, The total number of errors in the Z basis and X basis can be written as follows: 
\begin{widetext}
\begin{equation}
    \begin{split}
    m_{[\mu_a, \mu_b]}=n_{\text{tot}} \cdot (\frac{p_{\mu_a^e} p_{\mu_b^e} q_{(\mu_a^e|\mu_b^e)} p_{o_a^l} p_{o_b^l} q_{(o_a^l|o_b^l)}}{q_{\text{tot}}^2} 
                                     +\frac{p_{o_a^e} p_{o_b^e} q_{(o_a^e|o_b^e)} p_{\mu_a^l} p_{\mu_b^l} q_{(\mu_a^l|\mu_b^l)}}{q_{\text{tot}}^2}),
    \end{split}
\end{equation}
\begin{equation}
    \begin{split}
        m_{[2\nu_a, 2\nu_b]}=&\frac{n_{\text{tot}}}{M\pi}\cdot \int_0^{2\pi} \Big\{(1-e_d) 
                              \times \left[\frac{p_{\nu_a}^2 p_{\nu_b}^2 q_{(\nu_a|\nu_b)}^{\theta, L} q_{(\nu_a|\nu_b)}^{\theta+\delta, R}}{q_{\text{tot}}^2}+ 
                             \frac{p_{\nu_a}^2 p_{\nu_b}^2 q_{(\nu_a|\nu_b)}^{\theta, R} q_{(\nu_a|\nu_b)}^{\theta+\delta, L}}{q_{\text{tot}}^2}\right] \\  
                              &+e_d \cdot \left[\frac{p_{\nu_a}^2 p_{\nu_b}^2 q_{(\nu_a|\nu_b)}^{\theta, L} q_{(\nu_a|\nu_b)}^{\theta+\delta, L}}{q_{\text{tot}}^2} 
                            \frac{p_{\nu_a}^2 p_{\nu_b}^2 q_{(\nu_a|\nu_b)}^{\theta, R} q_{(\nu_a|\nu_b)}^{\theta+\delta, R}}{q_{\text{tot}}^2}\right] \Big\} d\theta,
    \end{split}
\end{equation}
\end{widetext}
Where $e_d$ represents the misalignment error rate, which can be found in Table \ref{t1}.
Then we could estimate the parameters we want.

(i) $\underline{s}_0^z$: $\underline{s}_0^z$ is the lower bound of the observed value of the total 
number of vacuum components in the Z basis, which means that Alice sends a vacuum state in the Z basis.
The lower bound of the expected value of the total number of vacuum components in the Z basis, $\underline{s}_0^{z*}$, could 
be expressed as~\cite{58:Zhou2023Experimental}: 
\begin{equation}
    \begin{split}
        \underline{s}_0^{z*}=\frac{e^{-\mu_a} p_{[\mu_a, \mu_b]}}{p_{[o_a, \mu_b]}} \underline{n}_{[o_a, \mu_b]}^{*},
    \label{B21}
    \end{split}
\end{equation}
where 
\begin{equation}
    \begin{split}
        p_{[k_a^{\text{tot}}, k_b^{\text{tot}}]}=\sum_{k_a^e+k_a^l=k_a^{\text{tot}}} \sum_{k_b^e+k_b^l=k_b^{\text{tot}}} {\frac{p_{k_a^e}p_{k_b^e}}{p_s} \frac{p_{k_a^l}p_{k_b^l}}{p_s}},
    \label{B22}
    \end{split}
\end{equation}
and 
\begin{equation}
    p_s=1-p_{\mu_a} p_{\nu_b}-p_{\nu_a} p_{\mu_b}.
    \label{B23}
\end{equation}

According to Eqs. (\ref{B21}) (\ref{B22}) (\ref{B23}) and (\ref{B2}), the lower bound of the observed value of the total number of vacuum components in the Z basis $\underline{s}_0^z=O^L(\underline{s}_0^{z*})$ could be obtained.

(ii) $\underline{s}_{11}^z$: $\underline{s}_{11}^z$ is the lower bound of the observed value of the number of 
single-photon pairs in the Z basis, which means that both Alice and Bob send a single-photon state in the Z basis.
The lower bound of the expected value of the number of single-photon pairs in the Z basis, $\underline{s}_{11}^{z*}$, 
could be expressed as~\cite{58:Zhou2023Experimental}: 
\begin{equation}
    \begin{split}
        \underline{s}_{11}^{z*} &\geq \frac{e^{-\mu_a-\mu_b} p_{[\mu_a, \mu_b]}}{\nu_a \nu_b (\mu'-\nu')} \\
                                &\times \left\{\mu_a \mu_b \mu' \left(e^{\nu_a+\nu_b} \frac{\underline{n}_{[\nu_a, \nu_b]}^{*}}{p_{[\nu_a,\nu_b]}} \right. \right. \\
                                &\left. \left. -e^{\nu_b} \frac{\overline{n}_{[o_a, \nu_b]}^{*}}{p_{[o_a,\nu_b]}}-e^{\nu_a} \frac{\overline{n}_{[\nu_a, o_b]}^{*}}{p_{[\nu_a,o_b]}}+\frac{\underline{n}_{[o_a, o_b]}^{*}}{p_{[o_a,o_b]}}\right)\right. \\
                                &\left. -\nu_a \nu_b \nu'  \left(e^{\mu_a+\mu_b} \frac{\overline{n}_{[\mu_a, \mu_b]}^{*}}{p_{[\mu_a,\mu_b]}} \right. \right. \\
                                &\left. \left. -e^{\mu_b} \frac{\underline{n}_{[o_a, \mu_b]}^{*}}{p_{[o_a,\mu_b]}}-e^{\mu_a} \frac{\underline{n}_{[\mu_a, o_b]}^{*}}{p_{[\mu_a,o_b]}}+\frac{\underline{n}_{[o_a, o_b]}^{*}}{p_{[o_a,o_b]}}\right)\right\}, 
    \end{split}
    \label{B24}
\end{equation}
where 
\begin{equation}
        \begin{split}
            &\mu'=\mu_a, \nu'=\nu_a \ \qquad \text{if} \ \frac{\mu_a}{\mu_b} \le \frac{\nu_a}{\nu_b} \\
            &\mu'=\mu_b, \nu'=\nu_b \ \qquad \text{if} \ \frac{\mu_a}{\mu_b} > \frac{\nu_a}{\nu_b}.
        \end{split}
    \label{B25}
\end{equation}

According to Eqs. (\ref{B24}) (\ref{B25}) (\ref{B22}) (\ref{B23}) and (\ref{B2}), the lower bound of the observed value of the total number of single-photon pairs in the Z basis $\underline{s}_{11}^z=O^L(\underline{s}_{11}^{z*})$ could be obtained.

(iii) $n_z$ and $E_z$: $n_z$ and $E_z$ each represents the length of the raw key without error correction and 
the bit error rate in the Z basis, which could be easily calculated through~\cite{58:Zhou2023Experimental} 
\begin{equation}
    n_z=n_{[\mu_a, \mu_b]},
\end{equation}
and
\begin{equation}
    E_z=\frac{m_{[\mu_a, \mu_b]}}{n_z},
\end{equation}
where $n_{[\mu_a, \mu_b]}$ represents the total number of bits in the Z basis, and $m_{[\mu_a, \mu_b]}$ represents the number 
of errors in the Z basis.

(iv) $\overline{\phi}_{11}^z$: $\overline{\phi}_{11}^z$ is the upper bound of the phase error rate in the Z basis, which 
could be estimated from $e_{11}^x$, the upper bound of the bit error rate of single-photon pair in the X basis. It could be expressed 
as~\cite{58:Zhou2023Experimental}: 
\begin{equation}
    \overline{e}_{11}^x=\frac{\overline{t}_{11}^x}{\underline{s}_{11}^x},
    \label{B28}
\end{equation}
in which $\overline{t}_{11}^x$ represents the upper bound of the observed value of the number of single-photon pair 
errors of the X basis and $\underline{s}_{11}^x$ represents the lower bound of the observed value of the number of 
single-photon pairs in the X basis. 

The lower bound of the expected value of the number of single-photon pairs in the X basis could be expressed 
as: 
\begin{equation}
    \begin{split}
    \underline{s}_{11}^{x*} &\ge \frac{e^{-2\nu_a-2\nu_b} 4 p_{[2\nu_a, 2\nu_b]}}{\mu_a \mu_b (\mu'-\nu')} \\
                            &\times \left\{\mu_a \mu_b \mu' \left(e^{\nu_a+\nu_b} \frac{\underline{n}_{[\nu_a, \nu_b]}^{*}}{p_{[\nu_a,\nu_b]}} \right. \right. \\
                            &\left. \left. -e^{\nu_b} \frac{\overline{n}_{[o_a, \nu_b]}^{*}}{p_{[o_a,\nu_b]}}-e^{\nu_a} \frac{\overline{n}_{[\nu_a, o_b]}^{*}}{p_{[\nu_a,o_b]}}+\frac{\underline{n}_{[o_a, o_b]}^{*}}{p_{[o_a,o_b]}}\right)\right. \\
                            &\left. -\nu_a \nu_b \nu'  \left(e^{\mu_a+\mu_b} \frac{\overline{n}_{[\mu_a, \mu_b]}^{*}}{p_{[\mu_a,\mu_b]}} \right. \right. \\
                            &\left. \left. -e^{\mu_b} \frac{\underline{n}_{[o_a, \mu_b]}^{*}}{p_{[o_a,\mu_b]}}-e^{\mu_a} \frac{\underline{n}_{[\mu_a, o_b]}^{*}}{p_{[\mu_a,o_b]}}+\frac{\underline{n}_{[o_a, o_b]}^{*}}{p_{[o_a,o_b]}}\right)\right\}. 
    \end{split}
    \label{B29}
\end{equation}

The upper bound of the number of single-photon pair errors of the X basis is: 
\begin{equation}
    \overline{t}_{11}^x \le m_{[2\nu_a, 2\nu_b]}-\underline{m}_{[2\nu_a, 2\nu_b]}^0,
    \label{B30}
\end{equation}
where
\begin{equation}
    \begin{split}
    \underline{m}_{[2\nu_a, 2\nu_b]}^{0*}&=e^{-2\nu_a} \frac{p_{[2\nu_a,2\nu_b]}}{2p{[o_a, 2\nu_b]}} \underline{n}_{[o_a, 2\nu_b]}^* \\
                                         &+e^{-2\nu_b} \frac{p_{[2\nu_a,2\nu_b]}}{2p{[2\nu_a, o_b]}} \underline{n}_{[2\nu_a, o_b]}^* \\
                                         &-e^{-2\nu_a-2\nu_b} \frac{p_{[2\nu_a,2\nu_b]}}{2p{[o_a, o_b]}} \overline{n}_{[o_a, o_b]}^*,
    \end{split}
    \label{B31}
\end{equation}
which represents the expected value of the lower bound of the error bit number in the X basis given that at least one of Alice and Bob 
sends a vacuum component.

Then we could get the upper bound of the bit error rate of single-photon pair in the X basis from Eqs. (\ref{B28})$-$(\ref{B31}) and (\ref{B2}).

Using the random sampling without a replacement theorem, with a failure probability $\epsilon_e$, we have the upper bound of a single-photon pair phase 
error rate in the Z basis\cite{58:Zhou2023Experimental} :
\begin{equation}
    \overline{\phi}_{11}^z \le \overline{e}_{11}^x+\gamma^U (\underline{s}_{11}^z, \underline{s}_{11}^x, \overline{e}_{11}^x, \epsilon_e).
    \label{B32}
\end{equation}

(v) $n$: Setting the length of signature $n$, the minimum length of $n$ that satisfies the security requirements, that is to say, satisfies Eq. (\ref{B34}), 
could be estimated with the calculated values of the parameters above by using the random sampling without replacement.\cite{82:Li2023One-time,91:yin2020tight}
The parameters in Eq. (\ref{B34}), $\underline{s}^{zn}_{0}$, the lower bound of vacuum events in a $n$-bit a 
selected key group, $\underline{s}^{zn}_{11}$, the lower bound of single-photon pairs events in the $n$-bit 
string, and $\underline{\phi}^{zn}_{11}$, the upper bound of the phase error rate of single-photon pairs in 
the $n$-bit string all need to satisfy~\cite{82:Li2023One-time}:
\begin{equation}
    \begin{split}
        \underline{s}_{0}^{zn}    & \ge n[\underline{s}_0^z/n_z - \gamma^U(n, n_z-n, \underline{s}_0^z/n_z, \epsilon)], \\
        \underline{s}^{zn}_{11}   & \ge n[\underline{s}_{11}^z/n_z - \gamma^U(n, n_z-n, \underline{s}_{11}^z/n_z, \epsilon)], \\
        \overline{\phi}_{11}^{zn} & \le \overline{\phi}_{11}^z + \gamma^U(\underline{s}_{11}^{zn}, \underline{s}_{zz}^z-\underline{s}_{11}^{zn}, \overline{\phi}_{11}^z, \epsilon). 
    \end{split}
    \label{B33}
\end{equation}

Then we have:
\begin{equation} 
    \mathcal{H}_n\leq \underline{s}_{0}^{zn}+\underline{s}^{zn}_{11}[1-H(\textcolor{black}{\overline{\phi}^{zn}_{11}})]-\lambda_{\rm EC},
    \label{B34}
\end{equation}
which represents the total unknown information of the $n$-bit string.

\section{SECURITY ANALYSIS}\label{Appendix3}
In order to disturb the authentication process, the attacker should try to make a difference in the results of the verification of Bob and Charlie~\cite{10:Yin2022Experimental}. Due to the existence of the leakage of information during the distribution stage, we divide this analysis into two parts. The first one takes the external attacker into account and the second one focuses on the QDS participants, mainly taking the internal attacker into account. 
\subsection{Attack from external attackers}
Unlike quantum key distribution that generates keys with perfect secrecy, in our protocol the keys are imperfectly secret. Any possible attackers may obtain partial information on the keys~\cite{82:Li2023One-time}. For the convenience of describing, we set the $m$-bits document $M$, then we could obtain that $\text{Sig}=h(M)\oplus r$, in which the function $h$ represents the hash function and the string $r$ represents the $Z_a$ in the description section. We could suppose the existence of an external attacker Eve, who has the ability to intercept and capture strings $\{\text{Sig},M\}$, tamper with it, and send it to the recipient, who will examine the signal he received before accepting it. 

Here we consider three types of attacks. The first one is to tamper the message randomly and relies entirely on fortune. The second one is to guess only $p_a$. The third one is to guess the keys from the captured signature. 

\subsubsection{\bf{Tampering randomly}}
We imagine a classical information $X$ of n-bits, and the attacker has access to a quantum system $E$ whose state $\rho^x_{E}$ depends on $X$. The attacker Eve can use $E$ to guess the string $X$ using an optimal strategy. We define $\mathcal{H}_n=H_{\text{min}}(X|E)_{\rho}$ as the min-entropy of $X$ and $E$, which can be estimated from the distribution stage~\cite{82:Li2023One-time}. According to the definition of min-entropy~\cite{92:Konig2009Operational}, we could get the probability of Eve correctly guessing $X$:
\begin{equation}
P_{\text{guess}}(X|E)=2^{-H_{\text{min}}(X|E)_{\rho}}=2^{-\mathcal{H}_n},
\end{equation}
and the $\mathcal{H}_n$ could be estimated from:
\begin{equation} 
    \mathcal{H}_n\leq \underline{s}^{zn}_{0}+\underline{s}^{zn}_{11}[1-H(\textcolor{black}{\overline{\phi}^{zn}_{11}})]-\lambda_{\rm EC},
    \label{C2}
\end{equation}
where $f$ is the error correction efficiency; $\underline{s}^{zn}_{0}$ is the 
lower bound of vacuum events in the $n$-bit string; $\underline{s}^{zn}_{11}$ is the lower bound 
of single-photon pairs events in the n-bit string; and $\underline{\phi}^{zn}_{11}$ represents 
the upper bound of the phase error rate of single-photon pairs in the n-bit string; $\lambda_{\rm EC}=nfH(E_{z})$ is the information consumed in the error correction stage of this string. All these parameters could be 
estimated from the distribution stage which is introduced in Appendix \ref{Appendix2}

After capturing $\{M,\text{Sig}\}$, what Eve should do is to tamper a new signal $\{M',\text{Sig}'\}$ and send it to the recipient, which will check that the signal satisfies $\text{Sig}'=h(M')\oplus r$ before accepting it. If the recipient accepts the $\{\text{Sig}',M'\}$, this attack will be deemed successful. The core point of the tamper is to make the $\text{Sig}'$ and $M'$ meet $\text{Sig}^{'}=h(M')\oplus r$, therefore, what the specific value of $\text{Sig}'$ or $M'$ is really does not matter so much. So, we can fix one of them and guess the other, and then the unknown information needing to be guessed is reduced to $n$ bits. So, for the first type of attack, $\mathcal{H}_n$ is equal to $n$. The success probability of this attack is:
\begin{equation}
P_1=2^{-n}.
\end{equation}
\subsubsection{\bf{Guessing keys}}
From the discussion above, we could know that the essence of attack is to guess the encryption method, in other words, the hash function in our method. The LFSR-based Toeplitz hash function we use can be expressed as:
\begin{equation}
h(M)=H_{nm}\cdot M.
\end{equation}
The crux of the function is the matrix $H_{nm}$, which is generated using $Y_a$ 
and $p_a$ in the messaging stage. From the Appendix \ref{Appendix1}, we could know that the attacker needs only to know $p_a$, so that 
Eve can easily generate a message $m$ of m-bits which satisfies $h(m)=0$, and the only 
requirement $m$ that needs to meet is $p_a(x)|m(x)$, in which $p_a(x)$ and $m(x)$ are polynomials 
generated from $p_a$ and $m$. We could get the success probability of this kind of 
attack~\cite{82:Li2023One-time}.
\begin{equation}
P_2=m\cdot2^{1-\mathcal{H}_n}=\epsilon_{\text{LFSR}}.
\end{equation}
We can obviously find that $P_2=\epsilon_{\text{LFSR}} \geq P_1$ in most occasions. 
\subsubsection{\bf{Recovering keys from the signature}}
This type of attack means that the attacker will try to recover the keys from the 
signature captured. In order to perform this kind of attack, the attacker needs to guess 
$Z_a$ and then perform the recovering algorithm. This will obviously lead to a smaller 
success probability compared to $\epsilon_{\text{LFSR}}$~\cite{82:Li2023One-time}. 
\subsection{Attack from internal attackers}
In this section we will put our attention on the QDS participants, considering the attackers from the internal, Alice or Bob. We don't consider Charlie as the attacker because he plays the role of notary. We divide this section into three sections, each considering one type of attack or error. 
\subsubsection{\bf{Robustness}}
This part will mainly consider the failure probability of the protocol when there are no attackers from the inside and outside. In other words, the three parties---Alice, Bob and Charlie---are all truthful. Therefore, the failure only occurs when Alice and Bob or Charlie share different keys after distribution stage, which will happen if there are some errors in the process of error correction or classical message transmission. 
We denoted this probability $\epsilon_{\text{rob}}=2\varepsilon_{\text{cor}}+2\varepsilon'$, in which 
$\varepsilon_{\text{cor}}$ and $\varepsilon'$ represents the error probability of  error correction 
and classical message transmission, respectively. 
\subsubsection{\bf{Repudiation}}
This kind of attack means that Alice wants to repudiate the established signature which was accepted by Bob, by making it rejected by Charlie, the notary. To make it accepted by Bob, there must be no error in distribution stage, so the only scenario in which repudiation succeeds is when there are errors existing in the process of the key exchange step. So the success probability can be expressed as $\epsilon_{\text{rep}}=2\varepsilon'$. 
\subsubsection{\bf{Forgery}}
In this attack, Bob will play the role of the attacker who wants to tamper with the message sent from Alice and send it to Charlie. Comparing this attack with external attacks, we could find that this attack is equal to the external attack where Bob plays the role of an external attacker. So we could get the success probability~\cite{82:Li2023One-time}:
\begin{equation}\label{C6}
\epsilon_{\text{for}}=m\cdot 2^{1-\mathcal{H}_n}
\end{equation}
\newline

From the discussion above, we see that the security bound of the scheme could be 
expressed as $\epsilon=\max\{\epsilon_{\text{rob}},\epsilon_{\text{rep}},\epsilon_{\text{for}}\}.$
\textcolor{black}{Above all, according to Eqs. (\ref{C6}) and (\ref{C2}), the security bound $\epsilon$ increases linearly as the document volume $m$ increases, but decreases exponentially as the unknown information of the potential attacker $\mathcal{H}_n$ increases.}

\section{SMOOTH MIN- AND MAX-ENTROPIES}\label{Appendix4}
The concept smooth min- and max-entropies is derived from the concept of min- and max-entropies, which is defined as below~\cite{92:Konig2009Operational}:
\begin{definition}\label{defD1}
    \textbf{Min-/Max-entropy:}
    Let $\rho=\rho_{AB}$ be a bipartite density operator. The min-entropy of $A$ conditioned on $B$ is defined by:
    \begin{equation}
        H_{\text{min}}(A|B):=-\underset{\sigma_B}{\text{inf}}D_{\infty}(\rho_{AB}||id_A\otimes\sigma_B),
    \end{equation}
    where the infimum ranges over all normalized density operators $\sigma_B$ on subsystem $B$ and where 
    \begin{equation}
    D_{\infty}(\tau||\tau'):=\text{inf}\{\lambda \in R :\tau\le 2^{\lambda}\tau'\}.
    \end{equation}
    The max-entropy is defined by:
    \begin{equation}
        H_{\text{max}}(A|B):=-H_{\text{min}}(A|C),
    \end{equation}
    where the min-entropy on the right-hand side is evaluated for a purification $\rho_{ABC}$ of $\rho_{AB}$. 
\end{definition}

Subsequently, we elucidate the definition of the smooth min- and max-entropies~\cite{92:Konig2009Operational}, \textcolor{black}{which is derived from min- and max-entropies for an optimal state $\rho^\prime$ in a $\varepsilon$-neighborhood of $\rho$.}
\begin{definition}\label{defD2}
    \textbf{Smooth Min-/Max-Entropy}
    Let $\rho=\rho_{AB}$ be a bipartite density operator and let $\varepsilon \ge 0$. The $\varepsilon$-smooth min- and max-entropies of $A$ conditioned on $B$ are given by:
    \begin{equation}
        H_{\text{min}}^{\varepsilon}(A|B)_{\rho}:=\underset{\rho'}{\text{sup}}H_{\text{min}}(A|B)_{\rho'},
    \end{equation}
    \begin{equation}
        H_{\text{max}}^{\varepsilon}(A|B)_{\rho}:=\underset{\rho'}{\text{inf}}H_{\text{max}}(A|B)_{\rho'},
    \end{equation}
    where the supremum ranges over all density operators $\rho'=\rho'_{AB}$ which are $\varepsilon$-close 
    to $\rho$.
\end{definition}

The smooth min- and max-entropies are closely related to quantum information and cryptography, which can 
help to analyze the length of the final key during the distribution through the theorems below~\cite{92:Konig2009Operational}:

\begin{theorem}\label{theD1}
Let $X$ be a classical random variable and let $B$ be (possibly quantum-mechanical) side information.
The smooth min-entropy is closely related to randomness extraction, which can, in the context of cryptography, turn a (only partially secure) 
raw key $X$ into a fully secure key $f(X)$ which is uniform and independent of the side information $B$~\cite{92:Konig2009Operational}. 

The maximum number of uniform and independent bits that can be extracted from $X$ is directly given by the smooth 
min-entropy of $X$. Let $l_{\mathrm{extr}}^{\varepsilon}(X|B)$ be the maximum length of a bit string that can be computed 
from $X$ such that $f(X)$ is $\varepsilon$-close to a string which is perfectly uniform and independent of the 
side information $B$. Then, the following connection exists:
\begin{equation}
    l_{\mathrm{extr}}^{\varepsilon}(X|B)=H_{\mathrm{min}}^{\varepsilon'}(X|B)+O(\mathrm{log}(1/\varepsilon)),
\end{equation}
where $\varepsilon' \in [\frac{1}{2}\varepsilon, 2\varepsilon]$.
\end{theorem}

\begin{theorem}\label{theD2}
    Considering a tripartite pure state $\ket{\Psi_{ABC}}$, the smooth max-entropy is closely related to 
    state merging, which aims to redistribute the $A$-part to the system $B$ by local operations and 
    classical communications (LOCC) between $A$ and $B$. Depending on the (reduced) state, this either 
    consumes or generates bipartite entanglement~\cite{92:Konig2009Operational}.
    
    Let $l_{\mathrm{merg}}^{\varepsilon}(A|B)_{\rho}$ be the minimal (maximal) number of ebits of entanglement 
    required (generated) by this process the distinction between consumed/generated entanglement is reflected by 
    the sign of the quantity $l_{\mathrm{merg}}^{\varepsilon}(A|B)_{\rho}$], such that the outcome is $\varepsilon$-close 
    to the desired output. Then, the following connection exists: 
    \begin{equation}
        l_{\mathrm{merg}}^{\varepsilon}(A|B)_{\rho}=H_{\mathrm{max}}^{\varepsilon'}(A|B)_{\rho}+O(\mathrm{log}1/\varepsilon),
    \end{equation}
    where $\varepsilon' \in [\frac{1}{2}\varepsilon, 2\varepsilon]$.
\end{theorem}

Supposing an eavesdropper Eve, we define $\bm{Z}$ as the raw key and $\bm{E}$ as the information of Eve learned from $\bm{Z}$ before error correction. We also define $\bm{Z}'$ as the key after error correction and $\bm{E}'$ as all information of Eve learned from $\bm{Z}$ after error correction. Let $\mathbb{H}$ denote the maximum length of a bit string that can be computed from $Z$ and $\varepsilon$-secure from the side information $E'$, i.e., $H_{\text{min}}^{\varepsilon}(\bm{Z}|\bm{E}')$, according to Theorem \ref{theD1}.
And we can easily get the expression of $H_{\text{min}}^{\varepsilon}(\bm{Z}|\bm{E}')$ in accordance with Definition \ref{defD1} \ref{defD2}, 
Theorem \ref{theD1} \ref{theD2} and the chain-rule inequality for smooth entropies~\cite{93:Vitanov2013Chain}:
\begin{equation}
    \begin{split}
        \mathbb{H} \ge H_{\text{min}}^{\varepsilon}(\bm{Z}|\bm{E})-H_{\text{max}}^{\varepsilon_{\text{cor}}}(\bm{Z}'|\bm{Z}).
    \end{split}
    \label{D8}
\end{equation}

Denote $H_{\text{min}}^{\varepsilon}(\bm{Z}|\bm{E})$ as $H_{\text{min}}^{\varepsilon}$ and $H_{\text{max}}^{\varepsilon_{\text{cor}}}(\bm{Z}'|\bm{Z})$ as $H_{\text{max}}^{\varepsilon_{\text{cor}}}$, then Eq.(\ref{D8}) could be simplified into Eq. (\ref{Eq2}) in Section \ref{Sec3}. 

Split $\bm{Z}$ into three parts: $\bm{Z}_0$, $\bm{Z}_{11}$ and $\bm{Z}_{\text{rest}}$, where $\bm{Z}_0$ s the bits where Alice sent a vacuum state, $\bm{Z}_{11}$ is the bits where both Alice and Bob sent a single photon and $\bm{Z}_{\text{rest}}$ is the rest of bits. 
Using a chain-rule for smooth entropies~\cite{93:Vitanov2013Chain}, we could get the expression:
\begin{equation}
    \begin{split}
    &H_{\text{min}}^{\varepsilon}(\bm{Z}|\bm{E}) \ge H_{\text{min}}^{\varepsilon'+2\varepsilon_e+(\hat{\varepsilon}+2\hat{\varepsilon}'+\hat{\varepsilon}^{\prime\prime})}(\bm{Z}_0\bm{Z}_{11}\bm{Z}_{\text{rest}}|\bm{E}) \\
                                                &\ge s_0^z+H_{\text{min}}^{\varepsilon_e}(\bm{Z}_{11}|\bm{Z}_0\bm{Z}_{\text{rest}}E)-2\text{log}_2 \frac{2}{\varepsilon'\hat{\varepsilon}},  
    \end{split}
    \label{D9}
\end{equation}
where $\varepsilon = \varepsilon'+2\varepsilon_e+(\hat{\varepsilon}+2\hat{\varepsilon}'+\hat{\varepsilon}^{\prime\prime})$.

Using the entropic uncertainty relation~\cite{94:Marco2011Uncertainty}, we have:
\begin{equation}
    \begin{split}
    H_{\text{min}}^{\varepsilon_e}(\bm{Z}_{11}|\bm{Z}_0\bm{Z}_{\text{rest}}E) &\ge s_{11}^z-H_{\text{max}}^{\varepsilon_e}(\bm{X}_{11}|\bm{X}'_{11})  \\
                                                                              &\ge s_{11}^z[1-H(\phi_{11}^z)].
    \end{split}
    \label{D10}
\end{equation}

According to Eqs. (\ref{D9}) (\ref{D10}), we could get Eq. (\ref{Eq3}).
Furthermore, the amount of bit information consumed during the error correction step could be expressed as:
\begin{equation}
    \begin{split}
    H_{\text{max}}^{\varepsilon_{\text{cor}}}(\bm{Z}'|\bm{Z}) &=\lambda_{EC}+\text{log}_2(\frac{2}{\varepsilon_{\text{cor}}}) \\
                            &=n_zfH(E_z)+\text{log}_2(\frac{2}{\varepsilon_{\text{cor}}}),\\&
    \end{split}
    \label{D11}
\end{equation}
where $f$ is the error correction efficiency. It can be rewritten as Eq. (\ref{Eq4}).

According to Eqs. (\ref{D8})-(\ref{D11}), we have:
\begin{equation}
    \begin{split}
        \mathbb{H} &= H_{\text{min}}^{\varepsilon}(\bm{Z}|\bm{E}')\\
                    &\ge s_{0}^z+s_{11}^z[1-H(\phi_{11}^z)]-n_z f H(E_z) \\
                    &\quad -2\text{log}_2(\frac{2}{\varepsilon^{\prime}\hat{\varepsilon}})-\text{log}_2(\frac{2}{\varepsilon_{\text{cor}}}),
    \end{split}
\end{equation}
where $\varepsilon_{\text{sec}}=2(\varepsilon'+2\varepsilon_e+\hat{\varepsilon}+2\hat{\varepsilon}'+\hat{\varepsilon}^{\prime\prime})$.
Then, we could finally get Eq. (\ref{Eq6}) in Section \ref{Sec3}. 
\section{SIMULATION DETAILS OF MDI-QDS}\label{Appendix5}
In the MDI-QDS~\cite{62:Puthoor2016Measurement}, the KGP protocol used between Alice, Bob and Charlie is a four-intensity protocol~\cite{88:Jiang2021Higher}. We take the KGP between  Alice and Bob as an example, during which Alice and Bob send pulses of intensity $k_{a(b)}\in\{\mu_{a(b)}, \nu_{a(b)}, \omega_{a(b)}, o_{a(b)}\}$. Here we denote the number and error number of detection events where Alice selects $k_a$ and Bob selects $k_b$ in the Z(X) basis as $n_{k_ak_b}^{z(x)}$ and $m_{k_ak_b}^{z(x)}$. They can be given by:
\begin{widetext}
\begin{equation}
\begin{split}
n_{k_ak_b}^{z}=N p_{k_a} p_{k_b} (1-p_d)^2 &e^{-\frac{k_a \eta_a+k_b \eta_b}{2}} \left\{p_d\cdot [I_0(\sqrt{k_a \eta_a k_b \eta_b}-(1-p_d)e^{-\frac{k_a \eta_a+k_b \eta_b}{2}}] \right. \\
&\left. +[1-(1-p_d)e^{-\frac{k_a \eta_a}{2}}][1-(1-p_d)e^{-\frac{k_b \eta_b}{2}}]\right\},
\end{split}
\end{equation}
\begin{equation}
n_{k_a k_b}^{x}=N p_{k_a} p_{k_b} y_{k_a k_b}^2 [1+2y_{k_a k_b}^2-4y_{k_a k_b} I_0(\frac{\sqrt{k_a \eta_a k_b \eta_b}}{2})+I_0(\sqrt{k_a \eta_a k_b \eta_b})],
\end{equation}
\begin{equation}
m_{k_ak_b}^{z}=N p_{k_a} p_{k_b} p_d (1-p_d)^2 e^{-\frac{k_a \eta_a+k_b \eta_b}{2}} [I_0(\sqrt{k_a \eta_a k_b \eta_b}-(1-p_d)e^{-\frac{k_a \eta_a+k_b \eta_b}{2}}],
\end{equation}
\begin{equation}
m_{k_ak_b}^{x}=N p_{k_a} p_{k_b} y_{k_a k_b}^2 \left\{1+y_{k_a k_b}^2-2y_{k_a k_b} I_0(\frac{\sqrt{k_a \eta_a k_b \eta_b}}{2})+e_d[I_0(\sqrt{k_a \eta_a k_b \eta_b})-1]\right\},
\end{equation}
\end{widetext}
where 
\begin{equation}
y_{k_a k_b}=(1-p_d)\cdot e^{-\frac{\eta_d(\eta_a k_a+\eta_b k_b)}{2}},
\end{equation}
and $e_d=0.04$.

By using the decoy-state analysis and the double-scanning method~\cite{88:Jiang2021Higher}, we can get the parameters of MDI-KGP as follows:
\begin{equation}
\begin{split}
&\underline{n}_0^{z*}=\text{max} \left\{\frac{e^{-\mu_a} p_{\mu_a}}{p_{o_a}} \underline{n}^{z*}_{o_a \mu_b}, \frac{e^{-\mu_b} p_{\mu_b}}{p_{o_b}} \underline{n}^{z*}_{\mu_b o_a}\right\},\\
&\underline{n}_{11}^{z*}=\frac{\mu_a \mu_b e^{-\mu_a-\mu_b} p_{\mu_a} p_{\mu_b}}{\nu_a \nu_b \omega_a \omega_b (\omega'-\nu')}
\left(\underline{P}^{+*}-\overline{P}^{-*}
+\underline{\hat{M}}^*-\overline{\hat{H}}^*\right),\\
&\overline{t}_{11}^{x*}=\frac{p_{\nu_a} p_{\nu_b}}{\omega_a \omega_b \omega' e^{\nu_a+\nu_b}}\left(\hat{M}-\frac{\hat{H}}{2}\right),\\
&\overline{t}_{11}^{z*}=\frac{\mu_a \mu_b e^{-\mu_a-\mu_b} p_{\mu_a} p_{\mu_b}}{\nu_a \nu_b e^{-\nu_a-\nu_b} p_{\nu_a} p_{\nu_b}}\cdot \overline{t}_{11}^{x*},\\
&\overline{\phi}^z_{11}=\frac{\overline{t}_{11}^{x}}{\underline{n}_{11}^{z}},\\
&E_z=\frac{m^z_{\mu_a \mu_b}}{n^z_{\mu_a \mu_b}},
\end{split}
\end{equation}
in which 
\begin{equation}
\begin{split}
&\omega'=\omega_a, \nu'=\nu_a, \ \text{if} \frac{\omega_a}{\omega_b} \le \frac{\nu_a}{\nu_b}, \\
&\omega'=\omega_b, \nu'=\nu_b, \ \text{if} \frac{\omega_a}{\omega_b} > \frac{\nu_a}{\nu_b},
\end{split}
\end{equation}
and
\begin{equation}
\begin{split}
&P^{+*}=\omega_a \omega_b \omega' e^{\nu_a+\nu_b} \frac{(n_{\nu_a \nu_b}^x-m_{\nu_a \nu_b}^x)^*}{p_{\nu_a} p_{\nu_b}} \\
&\ \ +\nu_a \nu_b \nu' e^{\omega_a} \frac{n^{x*}_{\omega_a o_b}}{p_{\omega_a} p_{o_b}}+\nu_a \nu_b \nu' e^{\omega_b} \frac{n^{x*}_{o_a \omega_b}}{p_{o_a} p_{\omega_b}}\\
&P^{-*}=\nu_a \nu_b \nu' e^{\omega_a+\omega_b} \frac{n^{x*}_{\omega_a \omega_b}}{p_{\omega_a} p_{\omega_b}}+\nu_a \nu_b \nu' \frac{n^{x*}_{o_a o_b}}{p_{o_a} p_{o_b}},\\
&\hat{M}^*=\omega_a \omega_b \omega' e^{\nu_a+\nu_b} \frac{m^{x*}_{\nu_a \nu_b}}{p_{\nu_a} p_{\nu_b}},\\
&\hat{H}^*=\omega_a \omega_b \omega' \left(e^{\nu_b} \frac{n_{o_a \nu_b}^{x*}}{p_{o_a} p_{\nu_b}} +e^{\nu_a} \frac{n_{\nu_a o_b}^{x*}}{p_{\nu_a} p_{o_b}} - \frac{n_{o_a o_b}^{x*}}{p_{o_a} p_{o_b}} \right).
\end{split}
\end{equation}

During the distribution, we scan $(\hat{H},\hat{M})$ to make the shared keys as secure as possible through the following programming:
\begin{equation}
\text{min} \ \ \  R
\end{equation}
\begin{equation}
\begin{split}
\text{such that}\ \ \ \ &\underline{\hat{H}}\le \hat{H} \le\overline{\hat{H}},\\
&\underline{\hat{M}}\le \hat{M} \le\overline{\hat{M}},
\end{split}
\end{equation}
where 
\begin{equation}
    \begin{split}
        R=&\frac{1}{N} \left\{\underline{n}_0^z+\underline{n}_{11}^z\left[1-H(\overline{\phi}^z_{11})\right]-\lambda_{EC}\right.\\
        &\left.-\text{log}_2 \frac{2}{\varepsilon_{\text{cor}}}-2\text{log}_2 \frac{2}{\varepsilon' \hat{\varepsilon}}-2\text{log}_2 \frac{1}{2\varepsilon_{\text{PA}}}\right\}. 
    \end{split}
\end{equation}
and 
\begin{equation}
\lambda_{EC}=n^z_{\mu_a \mu_b}fH(E_z).
\end{equation}

Denote the total length of raw key $n^z_{\mu_a \mu_b}$ as $n_z$. Denote the length of the signature as $L$ and the document volume as $m$. The signature rate per pulse pair could be given by:
\begin{equation}
R_{sig}=\frac{n_z}{2 L m}.
\end{equation}
in which, the length L is restricted by the security bound as follows~\cite{62:Puthoor2016Measurement}:
\begin{equation}
P(\text{honest abort}) \le 2e^{-(s_a-\overline{E}_z)^2 L}
\end{equation}
\begin{equation}
P(\text{repudiation}) \le 2e^{-(\frac{s_a-s_v}{2})^2 L}
\end{equation}
\begin{equation}
P(\text{forge}) \le 2e^{-(p_E-s_v)^2 L}
\end{equation}
where 
\begin{equation}
s_a=\overline{E}_z+\frac{p_E-\overline{E}_z}{4},
\end{equation}
\begin{equation}
s_v=\overline{E}_z+\frac{3(p_E-\overline{E}_z)}{4},
\end{equation}
and $p_E$ could be derived from:
\begin{equation}
c_0+c_1 \left[1-H(\overline{\phi}^z_{11})\right]=H(p_E),
\end{equation}
where $c_0=\underline{n}_0^z/n_z$ and $c_1=\underline{n}_{11}^z/n_z$.



%

\end{document}